\documentclass[12pt,a4paper]{article}
\usepackage{jheppub_kim}
 \usepackage{slashed}
\usepackage{longtable}
\usepackage{epsfig}
\usepackage{amssymb,amsmath}
\usepackage{orcidlink}
\usepackage{amsthm}
\usepackage{graphicx}
\usepackage{graphics}
\usepackage[hang,nooneline,scriptsize]{subfigure}
\usepackage{subfigure}
\usepackage{array}

\begin{document}
  
\title{Hawking Temperature of Massive Charged Ads Black Hole: a Topological Treatment}
 
\author[a]{Surajit Mandal\orcidlink{0000-0002-9091-0539}}
\author[b]{Surajit Das\orcidlink{0000-0003-2994-6951}\footnote{Corresponding author}}

\affiliation[a]{Department of Physics, Jadavpur University, Kolkata 700032, West Bengal, India}  
\affiliation[b]{IUCAA Centre for Astronomy Research and Development (ICARD), Department of Physics, Coochbehar Panchanan Barma University, Vivekananda Street, Coochbehar, 736101, West Bengal, India}
 
\emailAdd{surajitmandalju@gmail.com}
\emailAdd{surajit.cbpbu20@gmail.com} 

\abstract{ In this work, we investigate the Hawking temperature of a charged Ads black hole (spherically symmetric) on the basis of a completely topological method introduced by Robson, Villari, and Biancalana (RVB). This topological method can give the exact Hawking temperature of the charged Ads black hole. We have also derived the Hawking temperature of a charged Ads black hole considering massive gravity. Due to the presence of mass term in the metric function of the charged Ads black hole in massive gravity, the effect of mass term can't be neglected when calculating the Hawking temperature. In massive gravity, the accurate Hawking temperature can be obtained by including an integral constant term, which can be derived from the standard definition.
}	
	\maketitle 

\section{Introduction}
It is well established that the Laser Interferometer Gravitational-Wave Observatory (LIGO) Scientific Collaboration and Virgo Collaboration have detected the gravitational waves from inspiraling neutron star binaries and merging binary black holes \cite{b1,b2,b3,b4,b5}, which in turn gives powerful evidence for the existence of black holes. From the classical point of view, a black hole is always treated as an extreme object and it can't radiate anything. Meanwhile, Hawking and Bekenstein found that a black hole can produce temperature $T$ as well as entropy $S$ \cite{b6,b7,b8}. Therefore, one can consider the black hole system as a thermodynamical system \cite{b9,b10,b11,b12}. The mapping between a strong gravitational system and thermodynamics was proposed by Jacobson \cite{b13}.

The topological properties of a black hole can be perceived by the  Euler characteristic $\chi$ (topological invariant) \cite{b8,b15,b16,b17}. Recently, Robson, Villari, and Biancalana \cite{b18} exposed that the Hawking temperature of a black hole is nearly related to its topology. As a consequence, for calculating the Hawking temperature, they proposed a powerful topological method linked to the Euler characteristic $\chi$. This topological method has been successfully applied to Schwarzschild-like or charged black holes \cite{b21}, Schwarzschild-de Sitter black holes \cite{b19}, and anti-de Sitter black holes \cite{b20}. Moreover, taking into consideration this method, the Hawking temperature of a charged rotating BTZ black hole can be accurately derived by Liu et al. \cite{b20a}. Recently, Xian et al. \cite{b20b} showed that Hawking temperature of the global monopole spacetime can be calculated via topological formula. Chen et al. \cite{b20c} have successfully applied the RVB method for calculating the Hawking temperatures of conventional black holes under $f(R)$ gravity. The hawking temperature of a topological black hole using RVB method has been studied in \cite{b20d}. Previous work has connected the Hawking temperature to the topological properties of black holes. In Ref. \cite{b20b}, authors have applied the RVB method to a non-spherical symmetrical black hole in four-dimensional spacetime in massive gravity. In the present work, we shall apply the topological method to a four-dimensional spherically symmetric charged Ads black hole in massive gravity \cite{g1, Hendi}.

After a detailed derivation, we find that the Hawking temperature of the Schwarzschild--Ads, charged and charged Ads black hole in general relativity is obtained exactly by applying the RVB method. However, the Hawking temperature corresponding to the charged Ads black hole in massive gravity can not be estimated until an integral constant is not taken into consideration, which is mainly produced from the massive term. Moreover, starting from a reduced 2-dimensional black hole metric, we clearly show that RVB method can perfectly reproduce the Hawking temperature.

This paper is organized as follows. In Section \ref{sec1}, we present a brief review of the RVB method. In Section \ref{sec2}, we utilize the RVB method to obtain the temperature of the Schwarzschild-Ads, charged and charged Ads black hole cases in general relativity. For the charged Ads black hole in massive gravity, its Hawking temperature is also reproduced in Section \ref{sec3}. We present the graphical behavior of Hawking temperature with respect to the event horizon by changing various parameters like massive parameter, charge, and cosmological constant in section \ref{sec4}. Finally, we make our final remarks in section \ref{sec5}.

\section{Review of RVB Method}\label{sec1}
Both the rotating black holes and stationary black holes have simple metrics due to the variety of black hole systems, so the temperature can be easily calculated. However, it is difficult to calculate the temperature of many special black holes and this difficulty arises due to the complex coordinate system. Therefore, the RVB method anticipates the Hawking temperature of a black hole to the Euler characteristic $\chi$; this is enormously useful in determining the Hawking temperature in any coordinate system.

Recently, Robson, Villari, and Biancalana (henceforth RVB) have proposed a powerful topological method to derive the Hawking temperature of a black hole. Their study shows that the Hawking temperature for the black hole can be derived topologically. It was found that this method is effective for any coordinate system. A black hole has a topological invariant and the corresponding topological invariant is the Euler characteristic $\chi$ \cite{b8,b15,b16,b17}. There are various black hole systems decorated by different kinds of complicated metrics. In order to study them we usually require highly nontrivial coordinate transformations. But nowadays the situation is a little bit different, the Hawking temperature of a black hole can be calculated via the topological method linked to the Euler characteristic $\chi$.

In the RVB method, one can derive the Hawking temperature using the Euclidean geometry of the 2-dimensional spacetime keeping the information of the higher-dimensional spacetime. For a 2-dimensional black hole, the Hawking temperature can be estimated by the RVB method \cite{b18,b19} as follows:
\begin{equation}\label{a}
T_{H}=\frac{\hbar c}{4\pi\chi k_{B}} \Sigma_{j^{\prime}\leq \chi}\int_{r_{h_{j^{\prime}}}}\sqrt{|g|} \mathcal{R}dr.
\end{equation}

where the parameters $k_{B}$, $\hbar$, and $c$ refer to the Boltzmann constant, the Planck constant, and the speed of light respectively. The metric determinant is denoted by $g$, $r_{h_{j^{\prime}}}$ is the $j^{\prime}$-th Killing horizon. In the present study, we consider the parameters $\hbar=c=k_{B}=1$. Here $\mathcal{R}$ stands for the Ricci scalar of the 2-dimensional space-time. The Euler characteristic of Euclidean geometry is represented by the parameter $\chi$ and it can give the number of the Killing horizons. The summation over the Killing horizons is denoted by  $\Sigma_{j^{\prime}\leq\chi}$.

The Euler characteristic $\chi$ of a black hole is related to the structure of the manifold and it is a topological invariant. Here, the black hole space-time which has an outer boundary (event horizon) can be interpreted as a compact manifold. In a closed $n$-dimensional manifold $M^{n}$ (n= even), $\chi$ can be defined as \cite{r7}
\begin{equation}\label{b}
\chi=\frac{2}{\operatorname{area}\left(S^{n}\right)} \int_{M^{n}} \sqrt{|g|}G d^{n}x ,
\end{equation}
where area $\left(S^{n}\right)$ and $G$ denotes the surface area and density of the unit $n$-dimensional sphere, respectively. The density $G$ can be defined by using the Riemann coordinates as
\begin{equation}\label{c}
G=\frac{1}{2^{n / 2} n ! g} \epsilon^{i_{1} \cdots i_{n}} \epsilon^{j_{1} \cdots j_{n}} \mathcal{R}_{i_{1} i_{2} j_{1} j_{2}} \mathcal{R}_{i_{3} i_{4} j_{3} j_{4}} \cdots \mathcal{R}_{i_{n-1} i_{n} j_{n-1} j_{n}} ,
\end{equation}
where $\epsilon^{i_{1} \cdots i_{n}}$ stands for the Levi-Civita symbol in four-dimensions and $R_{\mu\nu\rho\tau}$ is Riemann tensor. In 2-dimensions, the density satisfies $G=\frac{\mathcal{R}_{1212}}{g}=\frac{\mathcal{R}}{2}$. The Euler characteristic is only calculated
at the Killing horizon of the black hole and shown in Ref. \cite{b18}.

In a two-dimensional spacetime, the Euler characteristic is
\begin{equation}\label{d}
\chi=\int \sqrt{|g|} d^{2}x \frac{\mathcal{R}}{4 \pi}.
\end{equation}
With the help of the Wick rotation $t=i \tau$ and $\beta=\frac{1}{T}$, the Euler characteristic $\chi$ takes the form \cite{b18}
\begin{equation}\label{e}
\chi=\int_{0}^{\beta} d \tau \int_{r_{\mathrm{H}}} \sqrt{|g|} d r \frac{\mathcal{R}}{4 \pi}.
\end{equation}
Finally, $\chi$ and the Hawking temperature $T_{H}$ will follow the following relation
\begin{equation}\label{f}
\frac{1}{4 \pi T_{H}} \int_{r_{\mathrm{H}}} \sqrt{|g|} \mathcal{R} d r=\chi ,
\end{equation}
which is the source of Eq. (\ref{a}).

\section{ Hawking temperature of the charged Ads black hole in general relativity}\label{sec2}
Now, we will apply the RVB method, the topological formula to study the Hawking temperature of a charged Ads black hole. The metric
of a standard charged Ads black hole is \cite{g1,g2}
\begin{equation}\label{1}
ds^2=-f(r)dt^2+\frac{dr^2}{f(r)}+r^2d\Omega_{2}^2 ,
\end{equation}
where $d\Omega_{2}^2$ stands for the standard element on $S^2$. Here, $f(r)$ denotes the lapse function and reads
\begin{equation}\label{2}
f(r)=1-\frac{2m}{r}+\frac{q^2}{r^2}-\frac{1}{3}\Lambda r^2 .
\end{equation}
where the parameters $m$, $q$ and $\Lambda$ correspond to the mass, charge, and cosmological constant. Cosmological constant related to Ads scale length as $\Lambda=-\frac{3}{l^2}$.

\begin{figure}[ht]
\begin{center}
\includegraphics[width=0.6\linewidth]{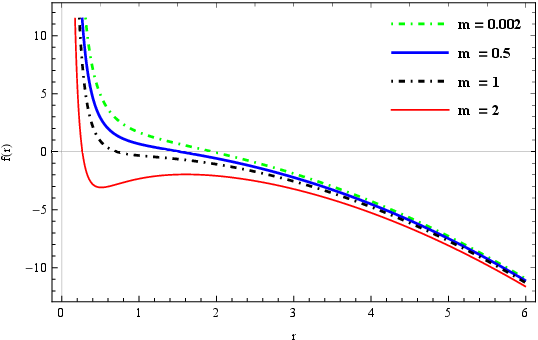}
\end{center}
\caption{ Plot of the lapse function $f(r)$ with resect to $r$ for different mass parameter $m$. Here, we set the parameters $q = 1, \Lambda = 1$.}
\label{fig1}
 \end{figure}
It is notified that the lapse function $f(r)$ depends on the charge and mass of the black hole. For different values of $m$ there is only one event horizon (Killing horizon) $r_{+}$, see Fig. \ref{fig1}. The Killing horizon shifted towards the higher $r$ for lower values of mass of charged Ads black hole.

Considering a special hypersurface, the charged Ads black hole can be streamed into a two-dimensional ($2D$) line element with a reduced metric through the Wick rotation ($\tau = it$) \cite{g3} as:
\begin{equation}\label{3}
ds^2=f(r)d\tau^2+\frac{dr^2}{f(r)} .
\end{equation}
Therefore, the Ricci scalar of the above reduced metric (\ref{3}) reads
\begin{equation}\label{4}
\mathcal{R}=-\frac{d^2 f(r)}{dr^2}=\frac{4m}{r^3}-\frac{6q^2}{r^4}-\frac{2}{3}\Lambda .
\end{equation}

\subsection{Schwarzschild-Ads black hole ($q=0$, $\Lambda\ne 0$)}\label{seca}
For the Schwarzschild-Ads black hole, the lapse function $f(r)$ is given in (\ref{2}) but for $q=0$, $\Lambda\ne 0$. The event horizon can be calculated when the lapse function $f(r)=0$. Interestingly, the computation reveals that among the three different roots of $f(r)$, only one is the real positive which in turn gives the radius of the horizon of the Schwarzschild-Ads black hole. The real positive root becomes
\begin{equation}\label{5}
r_{+}=-\frac{1}{\psi^\frac{1}{3}}-\frac{\psi^\frac{1}{3}}{\Lambda} ,
\end{equation}
where, $\psi=3m\Lambda^2+\sqrt{-\Lambda^3+9m^2\Lambda^4}$. Now, the Euler characteristic of the Schwarzschild--Ads black hole satisfies $\chi = 1$. Accordingly, the
Hawking temperature of this black hole can be calculated by using RVB method (\ref{a}):
\begin{equation}\label{6}
T_{H}=\frac{1}{4\pi r_{+}}\Big(1-\Lambda r_{+}^2\Big) ,
\end{equation}
which is consistent with the result calculated in Ref. \cite{g4}

\subsection{Charged black hole ($q\ne 0$, $\Lambda=0$)}\label{secb}
For the charged black hole, the lapse function $f(r)$ becomes
\begin{equation}\label{7}
f(r)=1-\frac{2m}{r}+\frac{q^2}{r^2} ,
\end{equation}
which is nothing but the lapse function of Reissner-Nordst\"{o}m black hole \cite{g4}. This charged black hole has two horizons
\begin{equation}\label{8}
r_{\pm}= m\pm\sqrt{m^2-q^2} .
\end{equation}
For special fixed values of parameters $m$ and $q$, there are two horizons $r_{-}$ (Cauchy horizon) and the $r_{+}$ (Killing horizon). There exist two horizon $r_{-}$ and $r_{+}$ for $q<m$, one degenerate horizon for $q=m$ and no horizon for $q>m$.

We notice that $f(r)$ has one killing horizon. Hence the Euler characteristic of the charged black hole is $\chi=1$. Plugging the Ricci scalar (\ref{4}) into Eq. (\ref{a}), the Hawking temperature
reads
\begin{equation}\label{9}
T_{H}=\frac{1}{4\pi r_{+}}\Big(1-\frac{q^2}{r_{+}^2}\Big) ,
\end{equation}
which is consistent with the standard result of Ref. \cite{g4,b18}
\subsection{Charged Ads black hole ($q\ne 0, \Lambda\ne0$)}\label{secc}
As discussed above, the charged Ads black hole has one horizon i.e. one Killing horizon $r_{+}$. The corresponding Euler characteristic of the charged Ads black hole is still $\chi=1$. In order to get Hawking temperature we use the RVB method and Hawking temperature can be calculated as
\begin{equation}\label{10}
T_{H}=\frac{1}{4\pi r_{+}}\Bigg[1-\Lambda r_{+}^2-\frac{q^2}{r_{+}^2}\Bigg].
\end{equation}
which is also consistent with the standard result of Refs. \cite{g1,g2}

\section{Hawking temperature of the charged Ads black hole in massive gravity}\label{sec3}
In this section, we will use the RVB method to estimate the Hawking temperature of the charged Ads black hole in massive gravity. First, we present a brief review of the charged Ads black hole in massive gravity. Using Vegh's approach for massive gravity, the action for $d$-dimensional massive gravity with the negative cosmological constant and Maxwell field in the presence of Maxwell source is \cite{Hendi}
\begin{eqnarray}\label{11}
{\mathcal A}=-\frac{1}{16\pi}\int d^dx \sqrt{-g}\left[R-2\Lambda+L(F)+\tilde{m}^2\sum_i^4c_iU_i(g,f)\right],\label{act}
\end{eqnarray}
where $R$ is a Ricci scalar, $\tilde{m}$ is the massive parameter, $\Lambda=-\frac{(d-1)(d-2)}{l^2}$ is a cosmological constant and $f$ refers to a fixed symmetric tensor. Here, $c_i$'s are constants and $U_i$'s are symmetric polynomials  \cite{g6}.

Here, we are interested in studying Maxwell electrodynamics, as a result, the function $L(F )$ is
\begin{equation}\label{12}
L(F)=-F ,
\end{equation}
where $F =F_{\mu\nu}F^{\mu\nu}$ (in which $F_{\mu\nu} = \partial{_{\mu}} A_{\nu} - \partial{_{\nu}} A_{\mu} $) stands for the electromagnetic field tensor and $A_{\mu}$ is the gauge potential. By varying the action (\ref{11}) with respect to the metric tensor $g_{\mu\nu}$ and the electromagnetic field tensor $F_{\mu\nu}$, leads to following equations of motion :
\begin{eqnarray}
G_{\mu\nu}+\Lambda g_{\mu\mu} +\frac{1}{2}g_{\mu\nu}F-2F_{\mu\lambda}F^\lambda_\nu +\tilde{m}^2\xi_{\mu\nu}& =&0,\\
\partial_\mu(\sqrt{-g}F^{\mu\nu})& =&0\label{g},
\end{eqnarray}
where $G_{\mu\nu}$ represents the Einstein tensor and $\xi_{\mu\nu}$ is the massive term
with the following form :
\begin{eqnarray}
\xi_{\mu\nu}&=& -\frac{c_1}{2}(U_1g_{\mu\nu}-K_{\mu\nu}) -\frac{c_2}{2}(U_2g_{\mu\nu}-2U_1K_{\mu\nu}+2K_{\mu\nu}^2)-\frac{c_3}{2}(U_3g_{\mu\nu} -3U_2K_{\mu\nu} \nonumber\\
&+&6U_1K_{\mu\nu}^2 -6K_{\mu\nu}^3)-\frac{c_4}{2}(U_4g_{\mu\nu} -4U_3K_{\mu\nu} +12U_2K_{\mu\nu}^2 -24U_1K_{\mu\nu}^3+24K_{\mu\nu}^4).
\end{eqnarray} 
Now, we will introduce the 4-dimensional static charged black holes in the context of massive gravity with AdS asymptotes. In order to do so, we consider a metric of 4-dimensional spacetime
in the form :
\begin{equation}\label{13}
ds^2=-f(r)dt^2+\frac{dr^2}{f(r)}+r^2(d\theta^2+sin^2\theta d\phi^2),
\end{equation}
considering the following reference metric \cite{g6,g7}
\begin{equation}\label{14}
f_{\mu\nu}=diag(0,0,c^2,sin\theta)
\end{equation}
where $c$ is a positive constant. Taking into account the reference metric written in Eq. (\ref{14}) for 4-dimensional spacetime, $U_{i}$'s takes the following forms \cite{g6,g7}
$$U_{1}=\frac{2c}{r}, U_{2}=\frac{2c^2}{r^2},U_{3}=0, U_{4}=0$$
With the help of the gauge potential ansatz $A_{\mu} = h(r)\delta_{\nu}^{0}$ in electromagnetic equation (\ref{g}) and taking the metric (\ref{13}), the lapse function $f(r)$ is obtained in Refs. \cite{Hendi, g6,g7} as
\begin{equation}\label{15}
f(r)=1-\frac{m_{0}}{r}-\frac{\Lambda}{3}r^2+\frac{q^2}{r^2}+\tilde{m}^2\Big(\frac{cc_{1}}{2}r+c^2c_{2}\Big) ,
\end{equation}
where $m_{0}$ and $q$ are the total mass ($m=\frac{m_{0}}{2}$) of black holes and the integration constants related to the electrical charge, respectively.

The corresponding Ricci scalar for the massive charged Ads black hole is
\begin{equation}\label{16}
\mathcal{R}=-\frac{d^2}{dr^2}f(r)=\frac{4m}{r^3}-\frac{6q^2}{r^4}-\frac{2}{3}\Lambda.
\end{equation}
Since the mass term $\tilde{m}$ in Eq. (\ref{15}) has no contribution to the Ricci scalar $\mathcal{R}$ and it makes the above Ricci scalar to be exactly same as the case of the charged Ads black hole in general relativity. Therefore, with the help of RVB method given in Eq. (\ref{a}), the Hawking temperature of the charged Ads black hole in massive gravity can be obtained as,
\begin{equation}\label{17}
T_{H}=\frac{1}{4\pi r_{+}}\Bigg[1-\Lambda r_{+}^2-\frac{q^2}{r_{+}^2}\Bigg],
\end{equation}
which is exactly equal to the case of $T_{H}$ obtained in Eq. (\ref{10}). Moreover, taking into consideration the standard definition of \cite{g8}, the Hawking temperature of the charged black hole in massive gravity is given by,
\begin{equation}\label{18}
T_{H}=\frac{1}{4\pi r_{+}}\Bigg[1-\Lambda r_{+}^2-\frac{q^2}{r_{+}^2}+\frac{\tilde{m}^2cc_{1}r_{+}}{2}\Bigg].
\end{equation}
We notice that, comparing (\ref{17}) and (\ref{18}), the Hawking temperature of the massive charged Ads black hole derived by the RVB method is quite different from that calculated by the standard definition. It seems that the RVB method is not applicable for the massive charged Ads black hole. However, we should observe that the topological formula (\ref{a}) is basically an indefinite integral and the indefinite integral has an integral constant. Keeping this into consideration, one can add an integral constant $c=\frac{\tilde{m}^2cc_{1}}{8\pi}$ into Eq. (\ref{17}) and as a result we will get the exact Hawking temperature (\ref{18}) for the massive charged Ads black hole.

It is worth mentioning that we can't neglect the effect of the mass term $\tilde{m}$ on temperature and the temperature modified by the standard definition \cite{g8} should be larger. Moreover, as long as the RVB method is taken into consideration, it is essential to add an integral constant to the expression of the Hawking temperature of a charged Ads black hole in massive gravity calculated by the standard definition.

However,  we can make the following simple calculation to eliminate the integral constant problem. Substituting Eq. (\ref{16}) into Eq. (\ref{a}), we can obtain $T_{H}$ for a $2D$ hypersurface given in the $\tau-r$ plane as
\begin{eqnarray}
T_{H}&=&\frac{1}{4\pi\chi}\int_{r_{+}}\frac{d^2}{dr^2}f(r)dr\nonumber\\&=&\frac{1}{4\pi\chi}\int_{r_{+}}df'(r)dr\nonumber\\&=&\frac{f'(r_{+})}{4\pi\chi}.
\end{eqnarray}
By considering $\chi=1$, one can get the same outcome of the temperature as that of the standard 'Euclidean trick'. So, we can exactly simulate the Hawking temperature of a black hole for the reduced metric (\ref{3}) with due help of the RVB method. Hence it is the generalized result for the RVB method.

\section{Graphical analysis of Hawking temperature}\label{sec4}
\begin{figure}[h!]
\begin{center} 
 $\begin{array}{cccc}
\subfigure[]{\includegraphics[width=0.35\linewidth]{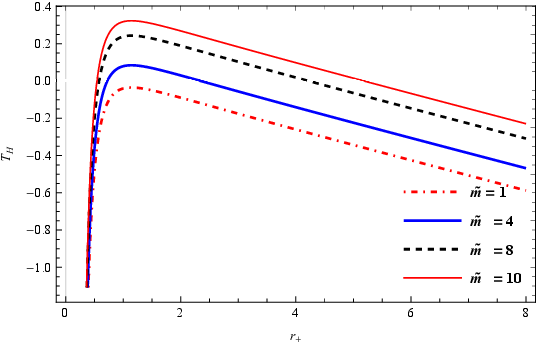}
\label{1a}}
\subfigure[]{\includegraphics[width=0.35\linewidth]{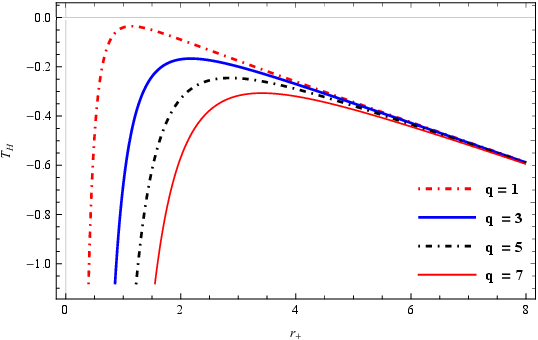}\label{1b}} 
\subfigure[]{\includegraphics[width=0.35\linewidth]{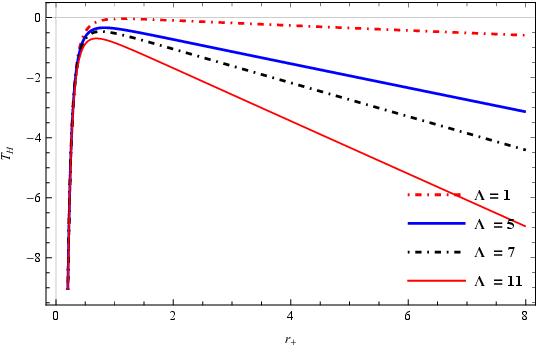}\label{1c}} 
\end{array}$
\end{center}
\caption{ Plot of $T_{H}$ with respect to event horizon radius $r_{+}$. We set the parameters $q=\Lambda=1$ [\ref{1a}], $\Lambda=\tilde{m}=1$ [\ref{1b}] and $\tilde{m}=q=1$ [\ref{1c}]. Here, we set $c=c_{1}=1$. }
\label{fig2}
\end{figure}

To show the behavior of Hawking temperature (Eq. \ref{18}) with event horizon radius, we plot Fig.\ref{fig2}. Figure \ref{1a} depicts that $T_{H}$ increases sharply for lower values of $r_{+}$. For higher values of $r_{+}$, the curve features decreasing nature. The value of $T_{H}$ increases for increasing massive parameter $\tilde{m}$ and remains positively valued. Figure \ref{1b} shows that $T_{H}$ increases sharply for lower $r_{+}$ but remains negatively valued and as like Fig.\ref{1a} the curve decreases for higher $r_{+}$. Here, $T_{H}$ decreases for increasing charge $q$.  However, Fig.\ref{1c} depicts that, $T_{H}$ decreases for increasing cosmological constant $\Lambda$.

\section{Summary and final remarks}\label{sec5}
In this section, We summarize our work. Firstly, we present a brief introduction and review of the topological RVB method.  By using the RVB method, the Hawking temperature of the Schwarzschild--Ads, charged and charged Ads black hole in general relativity can be perfectly obtained from the topological formula (\ref{a}). Here, we would like to mention that the Euler characteristic is defined in the even-dimensional manifold, and our study includes the four-dimensional charged Ads black hole which corresponds to the even-dimensional manifold. Even though our considered black hole is 4-dimensional, here in the present work, we can obtain the temperature of the black hole via the RVB method considering a 2-dimensional line element.

We also derived the Hawking temperature of the charged Ads black hole in massive gravity. Importantly, the direct use of the Ricci scalar (\ref{16}) to derive the Hawking temperature will lose the information given by the massive term $\tilde{m}^2(\frac{cc_{1}}{2}r+c^2c_{2})$. It resembles that the RVB method fails to give the exact temperature. Moreover, if one can add an integral constant term $c=\frac{\tilde{m}^2cc_{1}}{8\pi}$, then the correct Hawking temperature will be calculated. It is worth mentioning that, by plugging the Ricci scalar (\ref{16}) directly into the integral (\ref{a}), the Hawking temperature can be obtained properly. This indicates that the RVB method is an universal topological technique to extract the Hawking temperature for a black hole.

In future work, it wil be interesting to check whether the RVB method is applicable to higher-dimensional black holes and  also it would be interesting to consider the black hole which has cylindrical symmetry.

\end{document}